\RequirePackage{fix-cm}
\documentclass[twocolumn]{svjour3}          % twocolumn
\smartqed  % flush right qed marks, e.g. at end of proof
\usepackage{graphicx}
\usepackage{cite}
\usepackage{subfigure}
\usepackage{booktabs}
\usepackage{tabularx}
\usepackage{multirow}
\usepackage{stmaryrd}
\usepackage{amsmath}
\usepackage{verbatim}
\usepackage{url}

\newcommand{\argmax}{\operatornamewithlimits{argmax}}
\newcommand{\argmin}{\operatornamewithlimits{argmin}}
\newcommand{\etal}{\textit{et al.}~}

\journalname{Multimedia Systems}
\begin{document}

\title{Tag Relevance Fusion for Social Image Retrieval\thanks{The research was supported by 
NSFC (No. 61303184), 
SRFDP (No. 20130004120006), 
the Fundamental Research Funds for the Central Universities and the Research Funds of Renmin University of China (No. 14XNLQ01), 
and Shanghai Key Laboratory of Intelligent Information Processing, China (Grant No. IIPL-2014-002).}
}


\author{Xirong Li 
}


\institute{Xirong Li \at
              Key Lab of Data Engineering and Knowledge Engineering, Renmin University of China, 100872 China\\
              %No. 59 Zhongguancun Street, Beijing 100872, China, \\
              Shanghai Key Laboratory of Intelligent Information Processing, 200443 China \\
              %Tel.: +86-(0)10-62514562\\
              \email{xirong.li@gmail.com}           %  \\
}

\date{Received: date / Accepted: date}

\maketitle

\begin{abstract}

Due to the subjective nature of social tagging,
measuring the relevance of social tags with respect to the visual content is crucial for retrieving the increasing amounts of social-networked images.
Witnessing the limit of a single measurement of tag relevance,
we introduce in this paper tag relevance fusion as an extension to methods for tag relevance estimation. 
We present a systematic study, covering tag relevance fusion in early and late stages, and in supervised and unsupervised settings.
Experiments on a large present-day benchmark set show that tag relevance fusion leads to better image retrieval.
Moreover, unsupervised tag relevance fusion is found to be practically as effective as supervised tag relevance fusion,
but without the need of \emph{any} training efforts.
This finding suggests the potential of tag relevance fusion for real-world deployment.

\keywords{
Social image retrieval \and tag relevance estimation \and tag relevance fusion}
\end{abstract}

% ----------------------------------------------------------
\section{Introduction} \label{sec:intro}
% ----------------------------------------------------------

Searching for the ever growing amounts of varied and dynamically changing images on the social web is important for a number of applications.
The applications include
landmark visualization \cite{flickr-mm07},
visual query suggestion \cite{tomccap10-zha},
training data acquisition \cite{labelprop-tang-tist11},
photo-based question answering \cite{photoQA-mm08},
and photo-based advertisements \cite{advertise-mm08}, to name a few.
As users often assign tags when posting their images on social media,
one might expect tag-based retrieval to be a natural and good starting point for social image retrieval. %image search.
Compared to content-based search \cite{image-retrieval-survey08},
tag-based search potentially bypasses the semantic gap problem,
and its scalability has been verified by decades of text retrieval research \cite{modernIR}.
However, due to varied reasons, 
such as diversity in user knowledge, levels of expertise, and tagging intentions, 
social tagging is known to be ambiguous, subjective, and inaccurate \cite{Matusiak06Indexing}.
Moreover, since individual tags are used only once per image in the social tagging paradigm,
relevant and irrelevant tags for a specific image are not separable by tag statistics alone.
Measuring social tag relevance with respect to the visual content they are describing is essential.

For tag relevance estimation,
%for which
quite a few methods have been proposed.
%Quite a few methods have been proposed for social tag relevance estimation.
For example,
Liu \emph{et al}. propose a nonparametric method to rank tags for a given image by kernel density estimation in a specific visual feature space \cite{tagranking-www09}.
Chen \emph{et al}. train a Support Vector Machine classifier per tag \cite{tmm12-chen}. 
Given an image and its social tags,
Zhu \emph{et al.} propose to measure the relevance of a specific tag in terms of its semantic similarity to the other tags \cite{tmm12-zhu}.
In our earlier work \cite{tagrel-tmm09}, a neighbor voting algorithm is introduced which exploits tagging redundancies among multiple users.
By using learned tag relevance value as a new ranking criterion,
better image search results are obtained, when compared to image search using original tags.

Positioned in a deluge of social data, however, tag relevance estimation is challenging.
Visual concepts for example `boat or `garden' vary significantly in terms of their visual appearance and visual context.
%such as scale, viewpoint, and visual context.
%As a consequence,
A single measurement of tag relevance as proposed in previous work is limited %inadequate
to tackle such large variations, resulting in suboptimal image search.
At the feature level, it is now evident that no single feature
can represent the visual content completely \cite{smallgap-tmm10,VideoDiver-MIR07,Gehler-kernel-CVPR09,infosci14-lzhang,cyber14-lzhang}.
Global features are suited for capturing the gist of scenes \cite{gist-ijcv01},
while local features better depict properties of objects \cite{koen-tpami10,sp13-lzhang}.
As shown previously in content-based image search \cite{simplicity-pami01,wang:building},
image annotation \cite{AnnotationBaselines-IJCV10,TagProp-ICCV09},
and video concept detection \cite{mmfusion-mm04,graphfusion-csvt09},
fusing multiple visual features is beneficial.
So it is safe for us to envisage that tag relevance estimation will also benefit from the joint use of diverse features.
The question is \emph{what is the best strategy} to maximize such benefit?

Concerning fusion strategies, Snoek \emph{et al.} propose the taxonomy of early fusion and late fusion \cite{earlylatefusion-mm05},
which combine multiple sources of information at different stages.
\emph{Are early and late fusion schemes equally effective} for exploiting diverse features for measuring social tag relevance?
Moreover, for both schemes, supervised learning techniques have been developed to optimize fusion weights, see for instance \cite{wang:building,liu-learntorank}.
In principle, the learned weights, obtained at the cost of learning from many manually labeled examples, 
should be better than uniform weights which simply treats individual features (in early fusion) and individual tag relevance estimators (in late fusion) equally. 
However, this ``common sense'' is not necessarily valid for social media, which is large-scale, miscellaneous, and dynamically changes.
Towards coping with the many tags and many images in social media, it is worthy to ask: \emph{is supervised fusion a must?}

Towards answering the above questions, we make the following contributions:
%apt the notion of early and late fusion, and definewe ad

%
\begin{enumerate}
\item We propose visual tag relevance fusion as an extension of tag relevance estimation for social image retrieval.
      Using the neighbor voting algorithm as a base tag relevance estimator \cite{tagrel-tmm09},
      we present a systematic study on early and late tag relevance fusion.
      We extend the base estimator for both early and late fusion.
      Our previous work \cite{tagrel-civr10}, which discusses late tag relevance fusion only, is a special case of this work.
\item Experiments on a large benchmark \cite{nuswide-civr09} show that tag relevance fusion leads to better image search. 
      In particular, late fusion which combines both content-based \cite{tagrel-tmm09,tagranking-www09} and semantic-based \cite{tmm12-zhu} tag relevance estimators yields the best performance.
      Tag relevance fusion is also found to be helpful for acquiring better training examples from socially tagged data for visual concept learning.
\item This study offers a practical solution to exploit diverse visual features in estimating image tag relevance.
\end{enumerate}

The problem we study lies at the crossroads of social tag relevance estimation and visual fusion.
So next we present a short review of both areas.

%===============================================================
\section{Related Work} \label{sec:related-work}
%===============================================================

%===============================================================
\subsection{Social Tag Relevance Estimation} \label{ssec:related-tagrel}
%===============================================================

A number of methods have been proposed to attack the tag relevance estimation problem
\cite{tagranking-www09,tagrel-tmm09,tmm12-zhu,tmm12-chen,retagging-mm10,tagrepresent-mm10,tagrefine-icme10,tagrefine-mm10,tagrefine-lda-mm09}.
We structure them in terms of the main rationale they use, 
which is expressed in the following three forms, i.e., 
visual consistency \cite{tagranking-www09,tagrel-tmm09,tagrepresent-mm10,tagrefine-icme10}, semantic consistency \cite{tmm12-zhu},
and visual-semantic consistency \cite{retagging-mm10,tagrefine-mm10}.
%The rationale for  is that
Given two images labeled with the same tag,
the visual consistency based methods conjecture that if one image is visually closer to images labeled with the tag than the other image,
then the former image is more relevant to the tag.
Liu \etal \cite{tagranking-www09} employ kernel density estimation in a visual feature space to find such visually close images,
while Sun \etal exploit visual consistency to quantify the representativeness of an image with respect to a given tag \cite{tagrepresent-mm10}.
We introduce a neighbor voting algorithm which infers the relevance of a tag with respect to an image
by counting its visual neighbors labeled with that tag \cite{tagrel-tmm09}.
Lee \emph{et al}.  first identify tags which are suited for describing the visual content by a dictionary lookup \cite{tagrefine-icme10}.
Later, they apply the neighbor voting algorithm to the identified tags. 
To take into account negative examples of a tag which are ignored in the above works,
Chen \emph{et al}. train SVM models for individual tags \cite{tmm12-chen}. %, and use the models to estimate image tag relevance within the photo group. %to re-tag the photo group.
Li and Snoek take one step further by training SVM models with relevant positive and negative examples \cite{mm13-xli}.
Zhu \emph{et al}. investigate semantic consistency \cite{tmm12-zhu},
measuring the relevance of a tag to an image in terms of its semantic similarity to the other tags assigned to the image,
ignoring the visual content of the image itself.
Sun \emph{et al}. propose to use the position information of the tags, and tags appear top in the list are considered more relevant \cite{jasist11-sun}.
%assume that photos uploaded by the same user within a short time span form a semantically related group \cite{tmm12-chen} .
%They 
To jointly exploit visual and semantic consistency, % between tags,
Liu \etal
 perceive tag relevance estimation as a semi-supervised multi-label learning problem \cite{retagging-mm10},
while Zhu \etal formulate the problem as decomposing an image tag co-occurrence matrix \cite{tagrefine-mm10}.
Yang \emph{et al.} present a joint image tagging framework which simultaneously refines the noisy tags and learns image classifiers \cite{icmr2014-tagging-yang}.
Gao \etal propose to improve tag-based image search by visual-text joint hypergraph learning \cite{mm11-gao,tip13-gao}.
Given initial image search results, the authors view the top ranked images as positive instances, 
and re-rank the search results by hypergraph label propagation.
In all the above methods, only a single feature is considered. %used for tag relevance estimation.
%How to fuse multiple tag relevance estimates driven by diverse features remains largely unclear,
%which is the focus of this paper.
%We hypothesize that fusing multiple tag relevance estimates driven by diverse features
%could improve such methods.
%In all the above works, a single measurement of tag relevance is considered. 
How to effectively exploit diverse features for tag relevance estimation remains open. 
It is also unclear whether fusing the individual and heterogeneous measurements of tag relevance is beneficial.

%===============================================================
\subsection{Visual Fusion} \label{ssec:related-work-mmfusion}
%===============================================================

Snoek \emph{et al}. classify methods for visual fusion into two groups: early fusion and late fusion \cite{earlylatefusion-mm05}.
We follow their taxonomy to organize our literature review on visual fusion.
In early fusion, a straightforward method is to concatenate individual features
to form a new single feature \cite{earlylatefusion-mm05}.
As feature dimensionality increases, the method suffers from the curse of dimensionality \cite{hubness-jmlr10}.
Another disadvantage of the method is the difficulty to combine features into a common representation \cite{earlylatefusion-mm05}.
Instead of feature concatenation,
another method is to combine visual similarities of the individual features \cite{AnnotationBaselines-IJCV10,TagProp-ICCV09,graphfusion-csvt09}.
In these works, multiple visual (dis)similarities are linearly combined, with the combination weights optimized by distance metric learning techniques.
%In the context of image annotation, \cite{AnnotationBaselines-IJCV10} and \cite{TagProp-ICCV09}
%linearly combine multiple visual similarities.
In the context of video concept detection,  Wang \etal also chooses linear fusion to combine similarity graphs defined by different features \cite{graphfusion-csvt09}.
In a recent work for fine-grained image categorization \cite{tie14-lzhang},
an image is divided into multi-level hierarchical cells, and spatially adjacent cells are employed to describe the discriminative object components in a coarse-to-fine manner.
Graphlets are introduced in \cite{tip13-lzhang,tip14-lzhang} to describe multiple aspects of an image including spatial relationships between pixels and their color/texture distribution. 
In late fusion, models are obtained separately on the individual features
and their output is later combined \cite{mmfusion-mm04,VideoDiver-MIR07}.
In the work by Wu \etal \cite{mmfusion-mm04}, base classifiers are trained using distinct features, 
and the output of the base classifiers forms a new feature vector for obtaining a final classifier.
Wang \etal the base classifiers are combined in a boosting framework \cite{VideoDiver-MIR07}.
To the best of our knowledge, visual fusion in the tag relevance estimation context has not been well explored in the literature.

%===============================================================
\section{Base Tag Relevance Estimators} \label{sec:tagrel}
%===============================================================

For a valid comparison between early and late fusion,
we shall choose the same base tag relevance estimators for both fusion schemes.
Thus, before delving into the discussion about tag relevance fusion and its solutions,
we first make our choice of base estimators. % as follows.
For the ease of consistent description,
we use $x$ to denote an image, and $w$ for a social tag.
Let $g(x,w)$ be a base tag relevance function whose output is a confidence score of a tag being relevant to an image.
Further, let $\mathcal{S}$ be a source set of social-tagged images,
and $\mathcal{S}_{w}$ the set of images labeled with $w$, $\mathcal{S}_{w} \subset \mathcal{S}$.

A base estimator should be data-driven and favorably exploit the large amount of social data.
Moreover, it should be generic enough to adapt to both early and late fusion.
In that regard, we choose the neighbor voting algorithm proposed in our previous work \cite{tagrel-tmm09}.
Despite its simplicity, recent studies \cite{jasist11-sun,icme13-uricchio} report that this algorithm remains the state of the art for tag relevance estimation.
In order to find visual neighbors from $S$ for a given image $x$,
we use $z(x)$ to represent a specific visual feature vector.
We also have to specify a distance function for the given feature.
The optimal distance varies in terms of tasks \cite{zhang-survey2012}.
As the visual features used in this work, e.g., color correlogram and bag of visual words, are histogram based, we choose the $l_1$ distance. 
We use $\mathcal{S}_{x,z,k}$ to represent the $k$ nearest visual neighbors of $x$, retrieved by the $l_1$ distance on $z$.
The neighbor voting version of $g(x,w)$ is computed as
\begin{equation} \label{eq:neighbor-voting}
g(x,w) = \frac{|\mathcal{S}_{x,z,k} \cap \mathcal{S}_w|}{k} - \frac{|\mathcal{S}_w|}{|\mathcal{S}|},
\end{equation}
where $|\cdot|$ is the cardinality of a set.
The term $|\mathcal{S}_{x,z,k} \cap \mathcal{S}_w|$ is the number of neighbor images labeled with $w$.
Eq. (\ref{eq:neighbor-voting}) shows that more neighbor images labeled with the tag induces larger tag relevance scores,
and in the meantime, common tags with high frequency and thus less descriptive are suppressed by the second term.

\begin{figure*}[tb!]
\centering
 \subfigure[] {
\noindent\includegraphics[width=0.95\columnwidth]{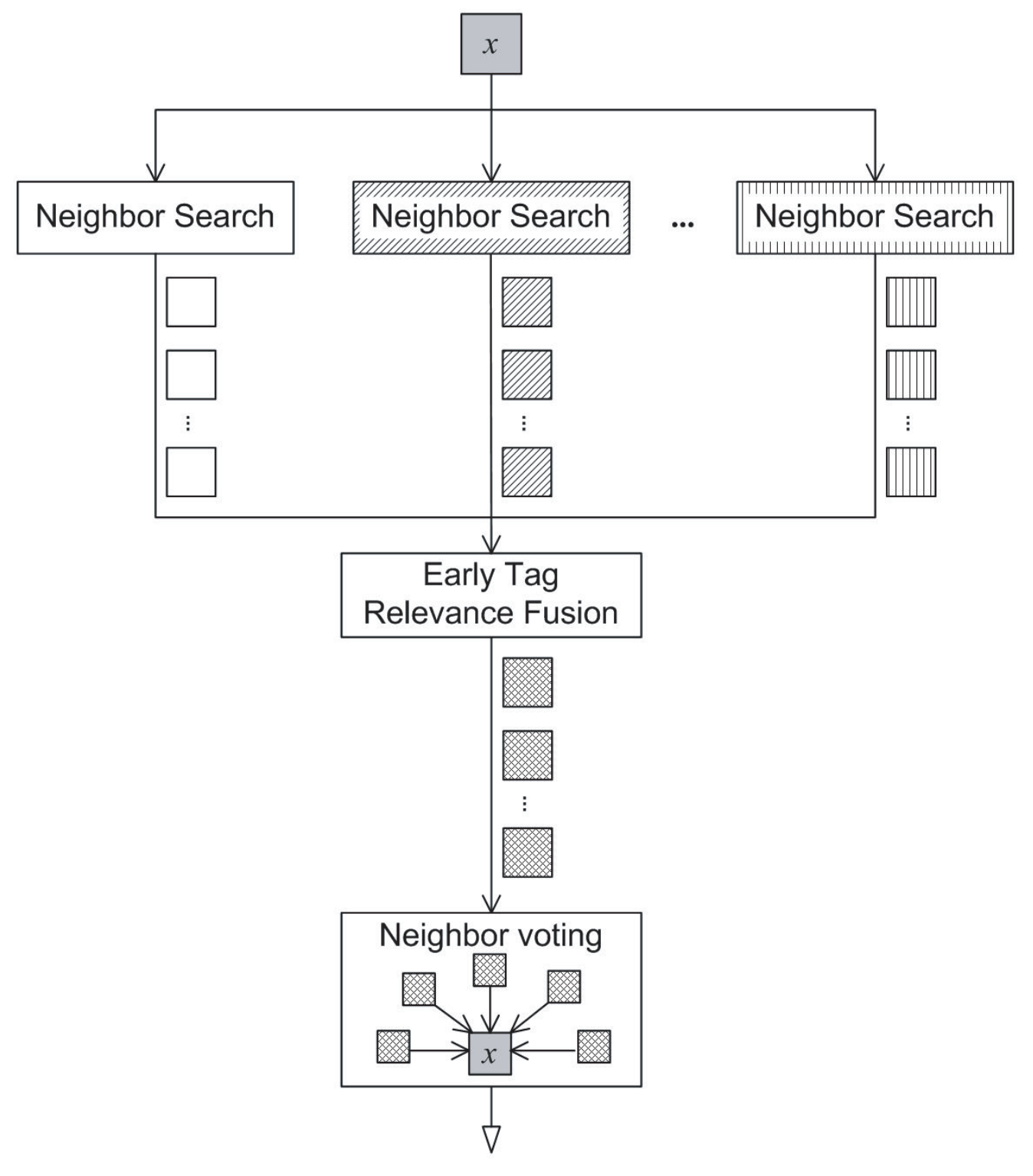}
\label{fig:earlyfusion}}
 \subfigure[] {
\noindent\includegraphics[width=0.95\columnwidth]{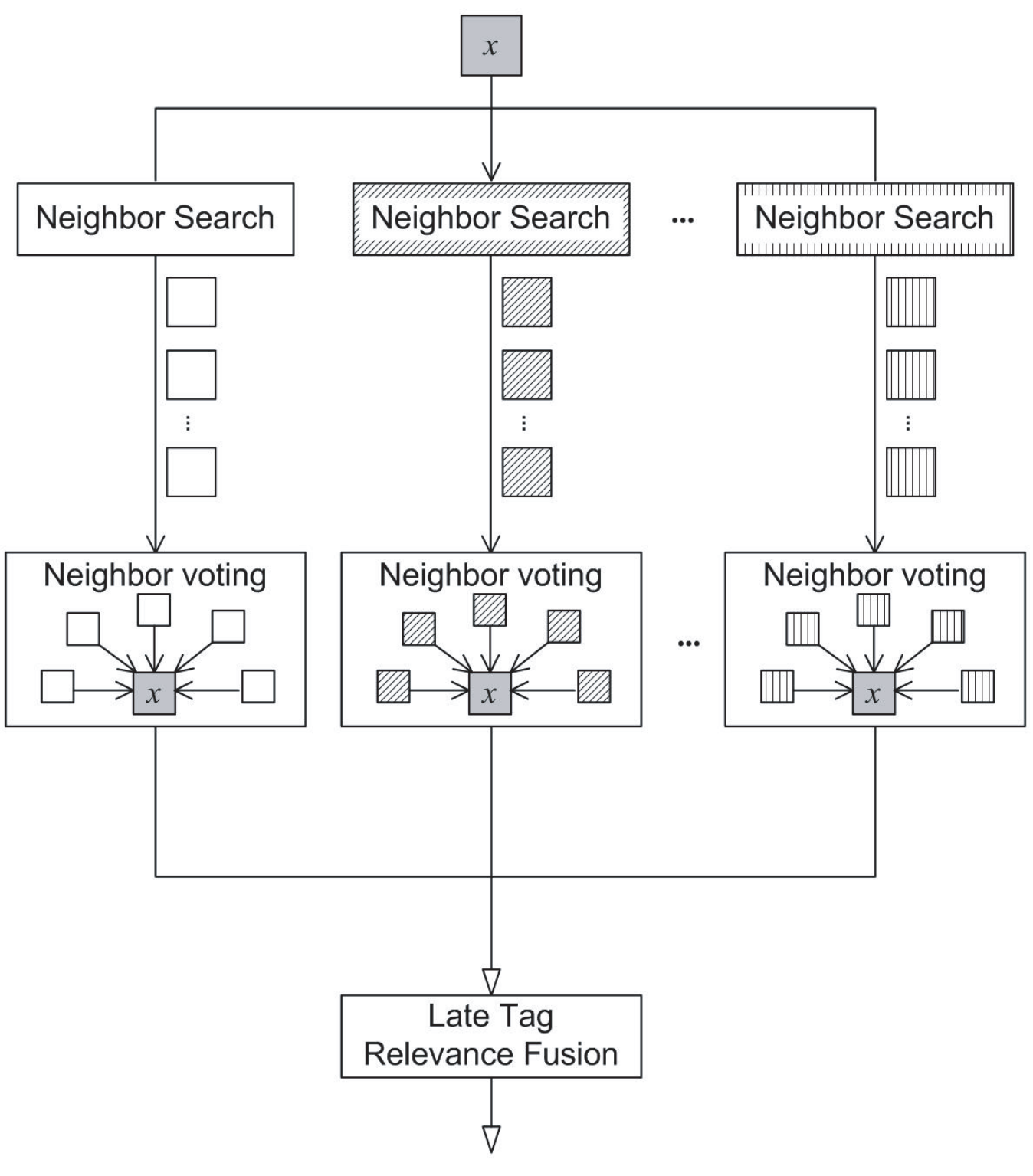}
\label{fig:latefusion}}
\caption{
\textbf{Extending the neighbor voting algorithm to (a) early tag relevance fusion and (b) late tag relevance fusion}.
Given an image $x$, different textured backgrounds indicate its visual neighbors obtained by distinct visual features.
In ealry tag relevance fusion, multiple visual neighbor sets are combined to obtain a better neighbor set for tag relevance estimation,
while in late tag relevance fusion, we fuse multiple tag relevance estimates.
} \label{fig:framework}
\end{figure*}

In what follows, we develop early and late fusion variants of the neighbor voting algorithm,
with a conceptual diagram illustrated Fig. \ref{fig:framework}.

%===============================================================
\section{Tag Relevance Fusion} \label{sec:tagrel-fusion}
%===============================================================

\subsection{Problem Formalization} \label{ssec:problem-formalization}

From an information fusion perspective \cite{diversity-fusion05}, diversity in base tag relevance estimators is important for effective fusion.
We generate multiple tag relevance estimators by
varying the visual feature $z$, the number of neighbors $k$, or both.
For a given feature, as a larger set of visual neighbors always include a smaller set of visual neighbors,
the parameter $k$ has a relatively limited impact on the diversity.
Hence, we fix $k$ and diversify the base estimators by using diverse visual features.
Let $\mathcal{Z}=\{z_1,\ldots,z_m\}$ be a set of such features, 
and $g_i(x,w)$ as a base estimator specified by feature $z_i$,
$i=1,\ldots,m$.
We adapt the notion of early and late fusion, defining

\medskip

\emph{Early Tag Relevance Fusion:} Fusion schemes that integrate individual features before estimating social tag relevance scores.

%\vspace{3pt}

\emph{Late Tag Relevance Fusion:} Fusion schemes that
first use individual features to estimate social tag relevance scores separately, and then integrate the scores.

\medskip

We use $G^e(x,w)$ to denote a fused tag relevance estimator obtained by early fusion,
and $G^l(x,w)$ to denote a late fused estimator. %for late fusion.
The goal of tag relevance fusion is to construct a $G(x,w)$, let it be $G^e(x,w)$ in early fusion 
and $G^l(x,w)$ in late fusion, so that when $G(x,w)$ is used as an image ranking criterion, 
better image retrieval is obtained compared to image retrieval using a single-feature estimator. % $g(x,w)$.

Since linear fusion is a well accepted choice for visual fusion as discussed in Section \ref{ssec:related-work-mmfusion},
we follow this convention for tag relevance fusion.
For early fusion,
we aim for a better neighbor set by combining visual similarities defined by the $m$ features.
Concretely, given two images $x$ and $x'$, let $d_i(x,x')$ be their visual distance computed in terms of feature $z_i$.
We define the combined distance as
\begin{equation} \label{eq:fuse-sim}
d_\Lambda(x,x')=\sum_{i=1}^m \lambda_i \cdot d_i(x,x'),
\end{equation}
where $\lambda_i$ is a weight indicating the importance of $z_i$.
The subscript $\Lambda$ is to make the dependence of the fused distance on $\{\lambda_i\}$ explicit.
We choose features which are intellectually devised,
so we assume that they are better than random guess, %We should choose take assume that the individual features are better than random guessing,
meaning adding them is helpful for measuring the visual similarity.
Hence, we constrain our solution with $\lambda_i \ge 0$.
Since normalizing weights by dividing by their sum does not affect image ranking,
any linear fusion with nonnegative weights can be transformed to a convex combination.
So we enforce $\sum_{i=1}^m \lambda_i=1$.
%We use $S_{x,\Lambda,k}$ to denote an image's $k$ nearest neighbors according to $sim_\Lambda(x,x')$.
%Replacing $S_{x,f,k}$ by $S_{x,\Lambda,k}$ in (\ref{eq:neighbor-voting}) leads to

Let $\mathcal{S}_{x,\Lambda,k}$ be the $k$ nearest neighbors retrieved by $d_\Lambda(x,x')$.
Substituting it for $\mathcal{S}_{x,z,k}$ in (\ref{eq:neighbor-voting}) 
leads to the early fused tag relevance function:
\begin{equation} \label{eq:early-fusion}
G^e_\Lambda(x,w)=\frac{|\mathcal{S}_{x,\Lambda,k} \cap \mathcal{S}_w|}{k} - \frac{|\mathcal{S}_w|}{|\mathcal{S}|}.
\end{equation}

In a similar fashion, we define the linear late fused tag relevance function:
\begin{equation} \label{eq:late-fusion}
G^l_\Lambda(x,w) = \displaystyle\sum_{i=1}^m \lambda_i \cdot g_i(x,w).
\end{equation}

\begin{comment}
By analyzing the statistical properties of $G^e_\Lambda(x,w)$ and $G^l_\Lambda(x,w)$ as given in the Appendix,
we show that the effectiveness of tag relevance estimation no longer counts on a specific feature, 
but on the majority of the features used. Hence, tag relevance fusion produces a more reliable measurement of image tag relevance.
\end{comment}

% ----------------------------------------------------------------------------
\subsection{Solutions for Tag Relevance Fusion} \label{ssec:fusion-methods}
% ----------------------------------------------------------------------------

As distinct features are of varied dimensions and scales, 
the resultant visual distance scores (and tag relevance scores) often reside at varied scales.
Score normalization is thus necessary before fusion.

% ----------------------------------------------------------------------------
\subsubsection{Score Normalization} \label{sssec:norm-methods}
% ============================================================================

We employ two popular strategies, i.e., MinMax and RankMax. 
Using a specific tag relevance estimator $g_i(x,w)$ as an example, its MinMax normalized version is defined as:
\begin{equation} \label{eq:minmax}
\tilde{g_i}(x,w)=\frac{g_i(x,w)-\min(g_i(x,w))}{\max(g_i(x,w))-\min(g_i(x,w))},
\end{equation}
where the min (max) function returns the minimum (maximum) possible score.
The RankMax normalized $g_i(x, w)$ is defined as:
\begin{equation} \label{eq:rankmax}
\hat{g_i}(x,w) = 1 - \frac{rank(g_i(x,w))}{n_w},
\end{equation}
where $rank(g_i(x,w))$ returns the rank of image $x$ when sorting images by $g_i(x,w)$ in descending order.
Compared to MinMax, RankMax quantizes scores into discrete ranks, making it more robust to outliers.

Intuitively, for early (late) tag relevance fusion, better features (estimators) should have larger weights.
Compared to the simplest solution that treats individual features and base estimators equally,
it is not surprising that when we have access to many well-labeled examples, a better solution can be learned.
However, for many tags, well-labeled examples are often of limited availability, 
making the study of unsupervised fusion necessary.
Therefore, we study tag relevance fusion in both unsupervised and supervised settings.

% ----------------------------------------------------------------------------
\subsubsection{Unsupervised Tag Relevance Fusion} \label{sssec:unsupervised-methods}
% ============================================================================

In an unsupervised setting, we have no prior knowledge of which feature or its resultant estimator is most appropriate for a given tag.
According to the principle of maximum entropy \cite{ProbabilityTheory-Jaynes-2003},
one shall make the least assumption about things we do not know. 
Hence, when no prior information concerning $\{\lambda_i\}$ is available, we shall use uniform weights.
Following this thought, we consider fusion by averaging.

\textbf{Unsupervised Early Fusion}.
%For early fusion, 
The fused distance $d_\Lambda(x,x')$ is the averaged value of $\{d_i(x,x')\}$, i.e., 
\begin{equation}
d_{avg}(x,x')=\frac{1}{m} \sum_{i=1}^m d_i(x,x').
\end{equation}

\textbf{Unsupervised Late Fusion}.
The corresponding $G^l_{\Lambda}(x,w)$ is simply the average of $\{g_i(x,w)\}$:
\begin{equation}
G^l_{avg}(x,w) = \frac{1}{m} \sum_{i=1}^m g_i(x,w).
\end{equation}
Notice that fusing the RankMax normalized functions with the uniform weights is equal to Borda Count,
a common algorithm for combining rankings generated by multiple sources of evidence \cite{Models-MetaSearch-SIGIR01}.

% ----------------------------------------------------------------------------
\subsubsection{Supervised Tag Relevance Fusion} \label{sssec:supervised-methods}
% ============================================================================

In an supervised setting, we aim to learn optimal fusion weights from many labeled examples.
For early tag relevance fusion, 
this is to optimize the combined distance so that the percentage of relevant neighbors will increase, 
and consequently better tag relevance estimation is achieved.
For late tag relevance fusion, this is to optimize the combined tag relevance estimator.
In the following, we describe two learning algorithms for the two fusion schemes, respectively.

\textbf{Supervised Early Fusion}.
%As discussed in Section \ref{sec:related-work}, 
Optimizing fusion weights at the distance level is essentially distance metric learning.
We opt to use the distance learning algorithm introduced by Wang \etal \cite{wang:building}, 
for its effectiveness for multi-feature neighbor search. % as well as its simplicity.
The basic idea is to find a combined distance 
to force images from the same class to be close, whilst images from different classes to be distant.
This is achieved by solving the following objective function:
\begin{equation} \label{eq:early-metric}
\argmin_{\Lambda} \sum_{x,x'}\left(\exp(-\sum_{i=1}^m \lambda_i \cdot d_i(x,x')) - y(x,x')\right)^2,
\end{equation}
where $(x,x')$ is a pair of images randomly sampled from the training data,
$y(x,x')=1$ if the two images have labels in common, and $y(x,x')=0$ otherwise.

\textbf{Supervised Late Fusion}.
Viewing the based estimators $\{g_i(x,w)\}$ as individual ranking criteria for image retrieval,
we tackle supervised late tag relevance fusion as a learning-to-rank problem. 
Let $E_{metric}(G^l_\Lambda(x,w))$ be a performance metric function which measures the effectiveness of $G^l_\Lambda(x,w)$ on a training set.
We seek $\Lambda$ that maximizes $E_{metric}$:
\begin{equation}\label{eq:coord}
\argmax_\Lambda E_{metric}(G^l_\Lambda(x,w)).
\end{equation}

Among many learning-to-rank algorithms, 
the coordinate ascent algorithm, developed by Metzler and Croft in the domain of document retrieval \cite{Metzler-LinearComination-IR07}, 
can directly optimize (non-differentiable) rank-based performance metrics, e.g., Average Precision and NDCG.
In the context of image auto-annotation \cite{ruc-clef2013}, 
we observe that weights learned by coordinate ascent consistently outperforms uniform weights for combining multiple meta classifiers. 
We therefore employ coordinate ascent for supervised late tag relevance fusion.

As a variant of hill climbing, coordinate ascent attempts to find $\Lambda$ that maximizes $E_{metric}$ in an iterative manner.
In each iteration, a better solution is found by changing a single element of the solution, i.e., the weight corresponding to a specific base estimator.
In particular, let $\lambda_i$ be the parameter being optimized.
We conduct a bi-direction line search with increasing steps to find the optimal value $\lambda^*_i$.
If the search succeeds, i.e., $\lambda^*_i$ yields a larger $E_{metric}$, we update $\lambda_i$ with $\lambda^*_i$.
Then, the next parameter $\lambda_{i+1}$ is activated, and the same procedure applies. 
The optimization process continues until the objective function no longer increases.

The two fusion schemes, combined with specific normalization and weighting methods,
result in the following 12 solutions:
\begin{enumerate}
\item Early-minmax-average: Early fusion with MinMax normalization and uniform weights;
\item Early-rankmax-average: Early fusion with RankMax normalization and uniform weights;
\item Early-minmax-learning: Early fusion with MinMax normalization and fusion weights optimized by distance metric learning;
\item Early-rankmax-learning: Early fusion with RankMax normalization and fusion weights optimized by distance metric learning;
\item Early-minmax-learning\textsuperscript{+}: Early fusion with MinMax normalization and fusion weights optimized per concept by distance metric learning;
\item Early-rankmax-learning\textsuperscript{+}: Early fusion with RankMax normalization and fusion weights optimized per concept by distance metric learning;
\item Late-minmax-average: Late fusion with MinMax normalization and uniform weights;
\item Late-rankmax-average: Late fusion with RankMax normalization and uniform weights;
\item Late-minmax-learning: Late fusion with MinMax normalization and fusion weights optimized by coordinate ascent;
\item Late-rankmax-learning: Late fusion with RankMax normalization and fusion weights optimized by coordinate ascent;
\item Late-minmax-learning\textsuperscript{+}: Late fusion with MinMax normalization and fusion weights optimized per concept by coordinate ascent;
\item Late-rankmax-learning\textsuperscript{+}: Late fusion with RankMax normalization and fusion weights optimized per concept by coordinate ascent.
\end{enumerate}

% ----------------------------------------------------------------------------
\subsection{Constructing Base Tag Relevance Estimators} \label{ssec:base-estimator}
% ----------------------------------------------------------------------------

As discussed in Section \ref{ssec:problem-formalization},
the parameter $k$ does not contribute significantly for diversifying the base estimators.
We empirically fix $k$ to be 500.
Concerning the features $\{z_i\}$, 
we choose the following four visual features which describe image content in different aspects: COLOR, CSLBP, GIST, and DSIFT.
COLOR is a 64-dimensional global feature \cite{color64-icme07}, combining a 44-d color correlogram,
a 14-d texture moments, and a 6-d RGB color moments.
CSLBP is a 80-d center-symmetric local binary pattern histogram \cite{cslbp-pr09}, 
capturing local texture distributions. 
GIST is a 960-d feature describing dominant spatial structures of a scene by a set of perceptual
measures such as naturalness, openness, and roughness \cite{gist-ijcv01}. 
DSIFT is a 1024-d bag of visual words depicting local information of the visual content,
obtained by quantizing densely sampled SIFT descriptors using a precomputed codebook of size 1,024 \cite{koen-tpami10}.
We will refer to the four base estimators using the corresponding feature names.

\section{Experimental Setup} \label{sec:setup}
% ----------------------------------------------------------

% ----------------------------------------------------------
\subsection{Data sets}  \label{ssec:dataets}
% ----------------------------------------------------------

\emph{Source set for constructing base estimators}.
To instantiate $\mathcal{S}$,
we use a public set of 3.5 million images\footnote{\url{http://pan.baidu.com/s/1gdd3dBH}} collected from Flickr in our previous work \cite{tagrel-tmm09}.
Since batch-tagged images tend to be visually redundant, we remove such images.
Also, we remove images having no tags corresponding to WordNet.
After this preprocessing step, we obtain a compact set of 815K images.

\emph{Benchmark data}. 
We choose NUS-WIDE \cite{nuswide-civr09}, a widely used benchmark set for social image retrieval.
This set contains over 250K Flickr images\footnote{\url{http://lms.comp.nus.edu.sg/research/NUS-WIDE.htm}
As some images are no longer available on Flickr, the dataset used in this paper are a bit smaller than the original release.
%While the NUSWIDE set originally contains around 260K images, a number of images are no longer available on Flickr.
}, 
with manually verified annotations for 81 tags which correspond to an array of objects, scenes, and event. 
As given in Table \ref{tab:data}, 
the NUS-WIDE set consists of two predefined subsets, one training set with 155,545 images and one testing set of 103,688 images.

%%%
% code to obtain numbers reported in this table
% python ~/Dropbox/myCode/cross-platform/tagprocess/countUser.py ~/VisualSearch/flickr81train/TextData/id.userid.lemmtags.txt 0
% python ~/Dropbox/myCode/cross-platform/tagprocess/countUser.py ~/VisualSearch/flickr81test/TextData/id.userid.lemmtags.txt 0
% python ~/Dropbox/myCode/cross-platform/tagprocess/wnvob.py flickr81test
%%%
\begin{table}[tb!]
\renewcommand{\arraystretch}{1.2}
\caption{\textbf{Data sets used in our experiments}.}
\label{tab:data}       % Give a unique label
\centering
\scalebox{1}{
\begin{tabular}{@{}l r r r@{}}
%\hline\noalign{\smallskip}
\toprule
& \textbf{Source set}  & \multicolumn{2}{c}{\textbf{NUS-WIDE}}  \\
\cmidrule{3-4}
&  & \textit{Training} & \textit{Test} \\
\cmidrule{1-4} % \cmidrule(lr){2-2} \cmidrule(lr){3-3}  \cmidrule(lr){4-4} \cmidrule(lr){5-5}
No. images              & 815,320 & 155,545 & 103,688 \\
No. users               & 177,871 &  40,202 & 32,415 \\
No. tags                &  34,429 &  28,367 & 25,278 \\
No. ground-truthed tags & N.A. & 81 & 81 \\
\bottomrule
\end{tabular}
}
\end{table}

% ----------------------------------------------------------
\subsection{Experiments}  \label{ssec:exp-retrieval}
% ----------------------------------------------------------

\subsubsection{Tag-based Image Retrieval}

We evaluate the effectiveness of tag relevance fusion in context of tag-based image retrieval.
That is, for each of the 81 test tags,
we sort images labeled with that tag in descending order by (fused) tag relevance scores.

\emph{Baselines}. As our goal is to study whether tag relevance fusion helps, 
the single-feature neighbor voting \cite{tagrel-tmm09} is a natural baseline.
For a more comprehensive comparison, 
we implement the following three present-day methods: 
tag position \cite{jasist11-sun}, 
tag ranking \cite{tagranking-www09},
and semantic field \cite{tmm12-zhu}.
As tag ranking requires a specific visual feature for kernel density estimation in the feature space,
we try tag ranking with each of the four features.

\emph{Evaluation Criteria}. 
We use Average Precision (AP), which is in wide use for evaluating visual search engines. 
We also report Normalized Discounted Cumulative Gain (NDCG), commonly used to assess the top few ranked results of web search engines \cite{ndcg}.
We compute NDCG for the top 100 ranked results.
For overall comparisons, we average AP and NDCG scores over concepts, reporting mAP and mNDCG.

\emph{Test of statistical significance}. 
We conduct significance tests, with the null hypothesis that there is no difference in mAP (or mNDCG) of two image retrieval systems.
In particular, we use the randomization test as recommended by Smucker \etal \cite{cikm07-msmucker}.

\subsubsection{Visual Concept Learning with Weak Labeling}

In this experiment, we apply tag relevance fusion to select better training examples for visual concept learning. 
The resultant concept classifiers will enable us to search images that are totally unlabeled.
Concretely, for each test tag, we select its positive training examples from the NUS-WIDE training set,
by sorting images in descending order by Late-minmax-average, and preserve the top 100 ranked images.
We consider SemanticField and TagRel$_{\text{COLOR}}$ as two baselines, 
applying them separately to acquire another two sets of 100 positive training examples.
As the focus is to compare which positive set is better, the same negative training data shall be used.
We take a random subset of 1,000 images from the NUS-WIDE training set as the common negative set, 
albeit more advanced methods for negative sampling exist \cite{tmm13-xli}. 
Fast intersection kernel SVMs \cite{cvpr08-fiksvm} are trained with the DSIFT feature, and later applied to classify the NUS-WIDE test set.

%Since frequent concepts such as people tend to have higher average precision scores than rare concepts like cow, 
%we also report a RandomGuess baseline for the ease of analysis, which is calculated as follows. 
%For each concept, we sort the test set at random and calculate mAP and mNDCG.  We run the process 100 times and take the averaged score.

% ----------------------------------------------------------
\section{Results} \label{sec:results}
% ----------------------------------------------------------

\subsection{Tag-Based Image Retrieval}

\begin{figure*}[tb!]
\centering
\noindent\includegraphics[width=2\columnwidth]{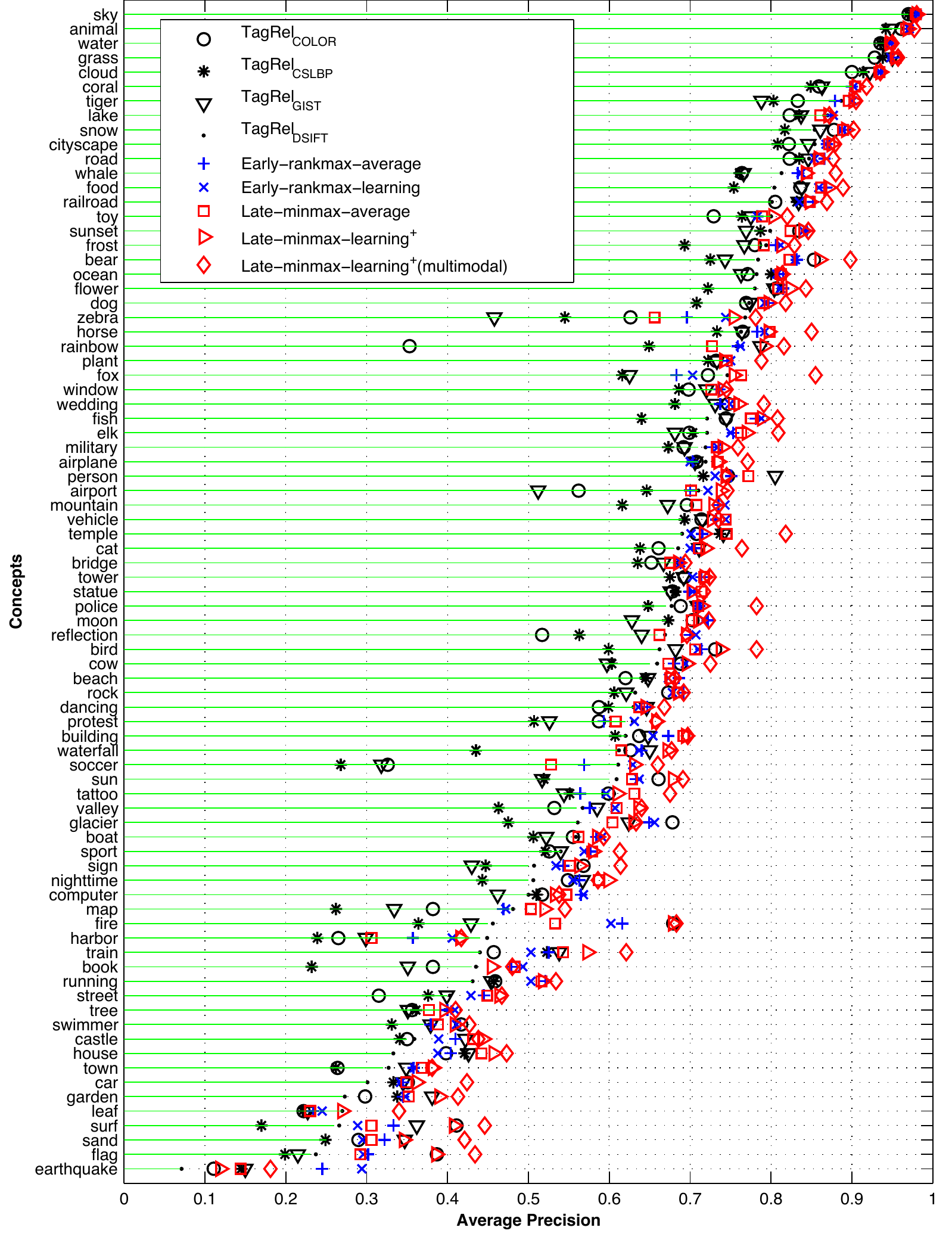}
\caption{\textbf{Tag Relevance Fusion versus Single Tag Relevance: A per-concept comparison}.
The concepts are sorted in descending order by TagRel$_{\text{DSIFT}}$. Best viewed in color.
}
\label{fig:per-concept}
\end{figure*}

\textbf{Tag relevance fusion \emph{versus} Single tag relevance}. 
As Table \ref{tab:overall-comparison} shows,
the best base estimator is TagRel$_{\text{DSIFT}}$, with mAP of 0.636 and mNDCG of 0.719.
Except for Early-minmax-average, 
all the other fusion solutions are significantly better than TagRel$_{\text{DSIFT}}$,
at the significance level of 0.01.
For a better understanding of the results, we make a per-concept comparison, see Fig. \ref{fig:per-concept}.
Compared to the best base estimator, 
tag relevance fusion improves AP scores for the majority of the concepts.
This can be observed from Fig. \ref{fig:per-concept} that the blue markers, representing early fusion,
and the red markers, representing late fusion, are mostly on the right side.
Further, for each concept we check the best performer among the four base estimators.
We find that for 21 concepts TagRel$_{\text{COLOR}}$ is the best, 
2 concepts for TagRel$_{\text{CSLBP}}$,
25 concepts for TagRel$_{\text{GIST}}$, and 34 concepts for TagRel$_{\text{DSIFT}}$.
Then, for every concept we compare Early-rankmax-average and Late-minmax-average with the concept's best performer,
which are concept dependent.
For 30 concepts, Early-rankmax-average outperforms the best performers, 
while Late-minmax-average beats the best performers for 46 concepts.
These results justify the effectiveness of visual fusion for improving tag relevance estimation.

%the neighbor voting algorithm beats tag position, semantic field, and tag ranking using different features.
%Tag relevance fusion leads to better image retrieval performance.
%For instance, Early-Average and Late-Average
%obtain an absolute improvement of 0.029 and 0.043, respectively,
%amounting to relative improvement of 4.6\% and 6.9\%.

\begin{table} [tb!]
\renewcommand{\arraystretch}{1.3}
\caption{\textbf{Performance of social image retrieval with and without tag relevance fusion}.
At the significance level of 0.01, the symbol * indicates that a fused tag relevance is better than the best single-feature 
tag relevance (TagRel$_{\text{DSIFT}}$), 
while the symbol \# indicates that a supervised fusion is better than its unsupervised counterpart.}
\label{tab:overall-comparison}
\centering
\scalebox{1}{
\begin{tabular}{@{}lll@{}}
\toprule

Method & mAP & mNDCG \\
\cmidrule{1-3}
\multicolumn{3}{@{}l}{\textbf{\emph{Baselines:}}} \\ 
TagPosition & 0.560   & 0.605\\
SemanticField & 0.577 & 0.607 \\ [3pt]
TagRanking$_{\text{COLOR}}$ & 0.578 & 0.596 \\
TagRanking$_{\text{CSLBP}}$ & 0.577 & 0.591 \\
TagRanking$_{\text{GIST}}$  & 0.575 & 0.589 \\
TagRanking$_{\text{DSIFT}}$ & 0.577 & 0.596 \\ [3pt]
TagRel$_{\text{COLOR}}$ & 0.625 & 0.712 \\
TagRel$_{\text{CSLBP}}$ & 0.588 & 0.657 \\
TagRel$_{\text{GIST}}$  & 0.621 & 0.710 \\
TagRel$_{\text{DSIFT}}$ & 0.636 & 0.719 \\ [3pt]
\multicolumn{3}{@{}l}{\textbf{\emph{Early tag relevance fusion:}}} \\ 
Early-minmax-average    & 0.646 & 0.734 \\
Early-rankmax-average      & 0.662\textsuperscript{*} & 0.756\textsuperscript{*} \\
%Early-Zero-Average      & ? & ? \\
Early-minmax-learning   & 0.657\textsuperscript{*,\#} & 0.749\textsuperscript{*,\#} \\ 
Early-rankmax-learning     & 0.664\textsuperscript{*}  & 0.755\textsuperscript{*} \\ % not applicable
Early-minmax-learning\textsuperscript{+}   & 0.658\textsuperscript{*,\#} & 0.749\textsuperscript{*,\#} \\ % not applicable
Early-rankmax-learning\textsuperscript{+}   & \textbf{0.665}\textsuperscript{*} & \textbf{0.756}\textsuperscript{*} \\ [3pt]
%Ealry-Zero-Learning     & ?  & ? \\ [3pt]
\multicolumn{3}{@{}l}{\textbf{\emph{Late tag relevance fusion:}}} \\ 
Late-minmax-average     & 0.660\textsuperscript{*} & 0.749\textsuperscript{*} \\
Late-rankmax-average       & 0.652\textsuperscript{*} & 0.739 \\
%Late-Zero-Average       &       &       \\
Late-minmax-learning    & 0.665\textsuperscript{*,\#} & 0.753\textsuperscript{*} \\
Late-rankmax-learning      & 0.659\textsuperscript{*,\#} & 0.745\textsuperscript{*} \\ % not applicable
%Late-Zero-Learning       &       &       \\
Late-minmax-learning\textsuperscript{+}  & \textbf{0.677}\textsuperscript{*,\#} & \textbf{0.773}\textsuperscript{*,\#} \\
Late-rankmax-learning\textsuperscript{+}  & 0.673\textsuperscript{*,\#} & 0.767\textsuperscript{*,\#}  \\ % not applicable
%Late-Zero-Learning\textsuperscript{+}   &  &  \\
\bottomrule
\end{tabular}
}
\end{table}

\textbf{Early tag relevance fusion \emph{versus} Late tag relevance fusion}.
%As shown in Table \ref{tab:overall-comparison},
There is no significant difference between early and late fusion in unsupervised settings.
%late fusion is in general better than early fusion, in both unsupervised and supervised settings.
Nevertheless, we observe the power of early fusion for addressing concepts that are rarely tagged.
Consider `earthquake' for instance. 
There are only 113 images labeled with the concept in $\mathcal{S}$.
The rare occurrence makes the base estimators mostly produce zero score for the concept.
Late fusion, with learning or not, does not add much in this case.
In contrast, by directly manipulating the neighbor sets, Early-rankmax-learning yields the best result for `earthquake'.
Notice that early fusion needs to combine tens of thousands of visual neighbors, making it computationally more expensive than late fusion.
Taking into account both effectiveness and efficiency, we recommend late fusion for tag relevance fusion.

For late fusion, Late-minmax-average, with mAP of 0.660 and mNDCG of 0.749, is slightly better than Late-rankmax-average, with mAP of 0.652 and 0.739.
For 54 concepts, Late-minmax-average outperforms Late-rankmax-average.
This result is mainly due to the fact that the base estimators already include an effect of smoothing by quantizing the visual neighborhood via neighbor voting.
Extra quantization by RankMax makes tag relevance estimates less discriminative.
Only when some base estimators yield large yet inaccurate values such as TagRel$_{\text{COLOR}}$ for `rainbow', Late-rankmax-average is preferred.

%In general, we consider LateFusion-Average the best choice for unsupervised fusion,
%for its competitive performance and its flexibility in adding new base estimators.

\textbf{Supervised fusion \emph{versus} Unsupervised fusion}.
The supervised methods achieve the best performance for both early and late fusion, see Table \ref{tab:overall-comparison}.
Supervised methods work particularly well for those concepts where there is large variance in the performance of the base estimators.
For early fusion, however, the difference between Early-rankmax-learning and Early-rankmax-average is not statistically significant.
For late fusion, the difference in mNDCG of Late-minmax-learning and Late-minmax-average is not statistically significant.
We also look into individual concepts.
Although for 49 concepts Late-minmax-learning improves over Late-minmax-average,
there are only 8 concepts having a relative improvement of more than 5\%.

Learning weights per concept is beneficial. 
For 65 concepts Late-minmax-learning\textsuperscript{+} is better than Late-minmax-average, 
and the number of concepts that have more than 5\% relative improvement increases from 8 to 17.
Nevertheless, because the weights are concept dependent, they are inapplicable to unseen concepts.

Overall, the performance of unsupervised fusion is close to supervised fusion. 
The result seems counter-intuitive as one would expect a larger improvement from supervised learning.
We attribute this to the following two reasons.
First, due to vagaries of social data,
for a number of concepts the models learned from the training data do not generalize well to unseen test data.
Second, different from traditional learning-to-rank scenarios where features or rankers might be just better than random guess \cite{liu-learntorank}, the features employed in this study were intellectually designed and shown to be effective.
As shown in Table \ref{tab:overall-comparison}, the base estimators already provide a strong starting point.
Moreover, distinct features result in complementary neighbor sets for early fusion and complementary tag relevance estimates for late fusion.
All this makes fusion with uniform weights a decent choice.
%Fusion with uniform weights is Therefore, though unsupervised fusion is suboptimal for some concepts, it is practically as effective as supervised fusion in general.

\textbf{Fusing heterogeneous tag relevance estimators}.
%As the base estimators are all on the base of visual neighbor voting, 
To study the effect of fusing heterogeneous tag relevance estimators,
we include semantic field and the four variants of tag ranking.
Comparing Table \ref{tab:overall-comparison} and Table \ref{tab:fuse-all}, 
we find that fusing the varied estimators is helpful.
Again, Late-minmax-average is comparable to Late-minmax-learning in terms of NDCG.
With mAP of 0.700 and mNDCG of 0.796, Late-minmax-learning\textsuperscript{+} performs best.
Note that the performance difference between Late-minmax-learning\textsuperscript{+} and Late-minmax-average becomes larger.
The result shows that concept-dependent weights are more needed for fusing tag relevance estimators driven by varied modalities.

\begin{figure*}[tb!]
\centering
\subfigure[Test tag: `military'] {
\noindent\includegraphics[width=2\columnwidth]{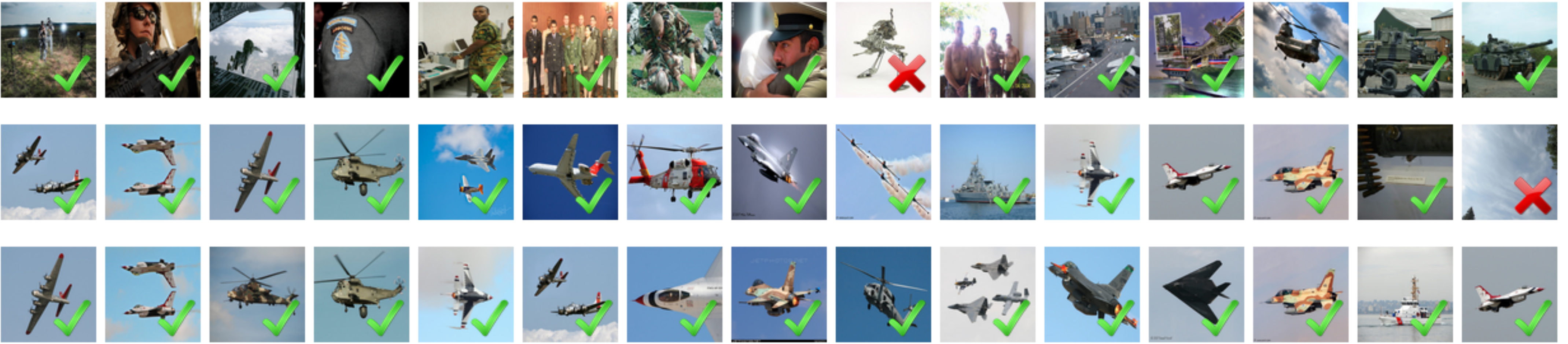}
\label{fig:top15-military}}
\subfigure[Test tag: `car'] {
\noindent\includegraphics[width=2\columnwidth]{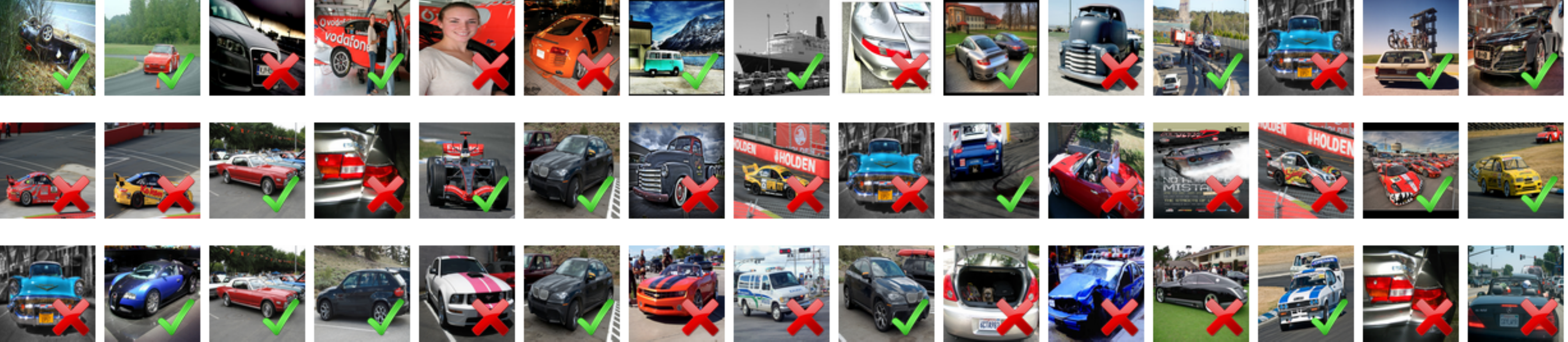}
\label{fig:top15-car}}
\subfigure[Test tag: `rainbow'] {
\noindent\includegraphics[width=2\columnwidth]{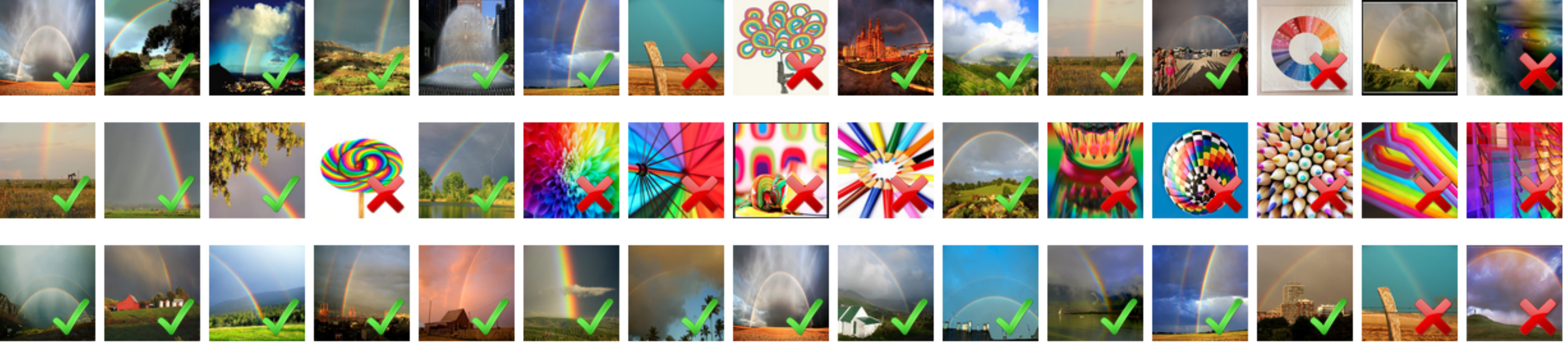}
\label{fig:top15-rawnbow}}
\subfigure[Test tag: `zebra'] {
\noindent\includegraphics[width=2\columnwidth]{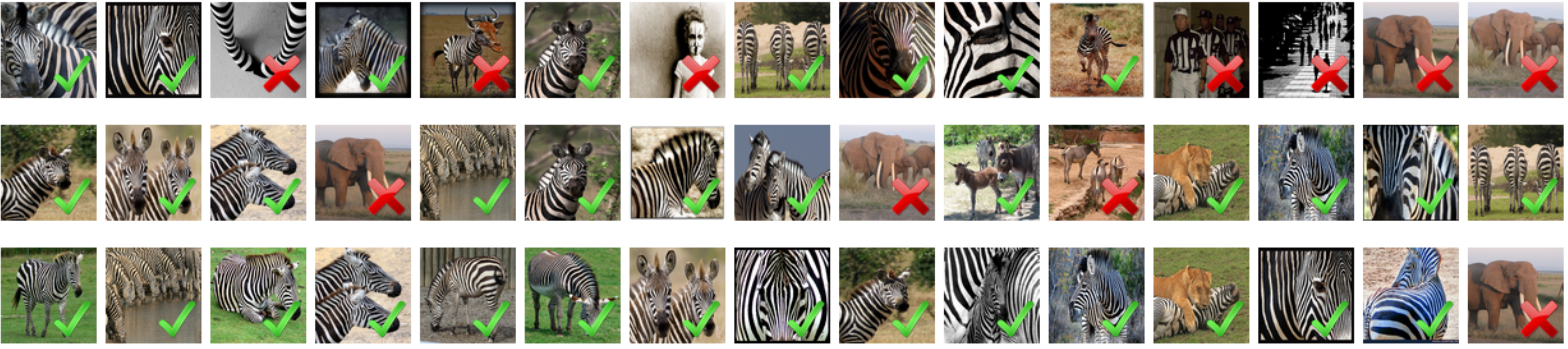}
\label{fig:top15-zebra}}
\caption{\textbf{Image retrieval results for test tags (a) `military', (b) `car', (c) `rainbow', and (d) `zebra'}.
From the top row to the bottom row, each subfigure shows the top 15 results returned by SemanticField \cite{tmm12-zhu}, TagRel$_{\text{COLOR}}$ \cite{tagrel-tmm09},
and the proposed Late-minmax-Learning\textsuperscript{+}, respectively.
Cross marks indicate false positives according to the NUS-WIDE annotation.
} \label{fig:results}
\end{figure*}

We present some image search results in Fig. \ref{fig:results}.
By exploiting diverse features,
tag relevance fusion is helpful for concepts having larger inter-concept visual ambiguity such as rainbow versus colorful things like balloons.
We observe from Fig. \ref{fig:top15-car} that the annotation of NUS-WIDE is incomplete:
a number of car images are not labeled as positive examples of `car'.
This is probably because the dataset developers used a kind of active learning strategy to ease the workload, without exhaustively labeling the dataset.

\begin{table} [tb!]
\renewcommand{\arraystretch}{1.3}
\caption{\textbf{Performance of tag-based image retrieval by fusing heterogeneous tag relevance estimators},
including the previous four base estimators, semantic field \cite{tmm12-zhu}, and four variants of tag ranking \cite{tagranking-www09}.
At the significance level of 0.01, the symbol \# indicates that a supervised fusion is better than its unsupervised counterpart.}
\label{tab:fuse-all}
\centering
\scalebox{1}{
\begin{tabular}{@{}lll@{}}
\toprule

Method & mAP & mNDCG \\
\cmidrule{1-3}
Late-minmax-average (multimodal)  & 0.673 & 0.759 \\
Late-minmax-learning (multimodal)                   & 0.679\textsuperscript{\#} & 0.763 \\
Late-minmax-learning\textsuperscript{+} (multimodal)  & 0.700\textsuperscript{\#} & 0.796\textsuperscript{\#} \\
\bottomrule
\end{tabular}
}
\end{table}

\subsection{Visual Concept Learning with Weak Labeling}

Table \ref{tab:concept-learning} shows the result of searching for the 81 test tags by the learned classifiers.
Notice that because the test set is treated as totally unlabeled in this experiment, 
the scores are much lower than their counterparts in Table \ref{tab:overall-comparison}.
We see from Table \ref{tab:concept-learning} that 
classifiers trained on positive examples selected by Late-minmax-average outperform classifiers trained on positive examples selected by the other methods.
Hence, tag relevance fusion is also helpful for acquiring better training examples for visual concept learning.

\begin{table} [tb!]
\renewcommand{\arraystretch}{1.3}
\caption{\textbf{Searching unlabeled images by visual concept classifiers learned from weakly labeled data}.
Classifiers trained on examples selected by Late-minmax-average beats 
classifiers trained on examples selected by the two baselines.}
\label{tab:concept-learning}
\centering
\scalebox{1}{
\begin{tabular}{@{}lrr@{}}
\toprule

 Positive example selection  & mAP & mNDCG \\
\cmidrule{1-3}
%RandomGuess             & & \\
SemanticField           & 0.119 & 0.271 \\
TagRel$_{\text{COLOR}}$ & 0.119 & 0.298 \\
Late-minmax-average     & \textbf{0.127} & \textbf{0.339} \\
\bottomrule
\end{tabular}
}
\end{table}

% ----------------------------------------------------------
\section{Discussion and Conclusions} \label{sec:conclus}
% ----------------------------------------------------------

Tag relevance estimation is important for social image retrieval.
On recognizing the limitations of a single measurement of tag relevance,
we promote in this paper tag relevance fusion as an extension to tag relevance estimation.
We develop early and late fusion schemes for a neighbor voting based tag relevance estimator,
and systematically study their characteristics and performance.
Image retrieval experiments on a popular benchmark set of 250K images justify our findings as follows.

1) Tag relevance fusion improves tag relevance estimation. 
Comparing to the four base estimators whose mAP scores ranging from 0.588 to 0.636, 
fused tag relevance results in higher mAP ranging from 0.646 to 0.677.
Adding extra heterogeneous estimators lifts mAP to 0.700.

2) The two fusion schemes each have their merit.
By directly manipulating the visual neighbors, early tag relevance fusion is more effective for addressing concepts that are rarely tagged.
Late fusion allows us to directly optimize image retrieval, and it is more flexible to handle varied tag relevance estimators.

3) Supervised fusion is meaningful only when one can afford per-concept optimization. 
Concept-independent weighting is marginally better than averaging the base estimators.
For tag relevance fusion, we recommend the use of Late-minmax-average as a practical strategy.

\medskip

\textbf{Acknowledgments}. 
The author is grateful to dr. Cees Snoek and dr. Marcel Worring for their comments and suggestions on this work.

\end{document}